\documentstyle[psfig]{mn}

\newif\ifAMStwofonts

\newcommand{\fmmm}[1]{\mbox{$#1$}}

\newcommand{\scnp}{\mbox{\fmmm{''}}}

\ifoldfss
  \ifCUPmtlplainloaded \else
    \NewTextAlphabet{textbfit} {cmbxti10} {}
    \NewTextAlphabet{textbfss} {cmssbx10} {}
    \NewMathAlphabet{mathbfit} {cmbxti10} {} 
    \NewMathAlphabet{mathbfss} {cmssbx10} {} 
  \fi
  \ifAMStwofonts
    \ifCUPmtlplainloaded \else
      \NewSymbolFont{upmath} {eurm10}
      \NewSymbolFont{AMSa} {msam10}
      \NewMathSymbol{\upi}     {0}{upmath}{19}
      \NewMathSymbol{\umu}     {0}{upmath}{16}
      \NewMathSymbol{\upartial}{0}{upmath}{40}
      \NewMathSymbol{\leqslant}{3}{AMSa}{36}
      \NewMathSymbol{\geqslant}{3}{AMSa}{3E}

      \let\leq=\leqslant \let\le=\leqslant
      \let\geq=\geqslant 
    \fi
  \fi
\fi 

\ifnfssone
  \newmathalphabet{\mathit}
  \addtoversion{normal}{\mathit}{cmr}{m}{it}
  \addtoversion{bold}{\mathit}{cmr}{bx}{it}
  \newmathalphabet{\mathbfit} 
  \addtoversion{normal}{\mathbfit}{cmr}{bx}{it}
  \addtoversion{bold}{\mathbfit}{cmr}{bx}{it}
  \newmathalphabet{\mathbfss} 
  \addtoversion{normal}{\mathbfss}{cmss}{bx}{n}
  \addtoversion{bold}{\mathbfss}{cmss}{bx}{n}
  \ifAMStwofonts
    \ifCUPmtlplainloaded \else
      \UseAMStwoboldmath
      \makeatletter
      \new@mathgroup\upmath@group
      \define@mathgroup\mv@normal\upmath@group{eur}{m}{n}
      \define@mathgroup\mv@bold\upmath@group{eur}{b}{n}
      \edef\UPM{\hexnumber\upmath@group}
      \new@mathgroup\amsa@group
      \define@mathgroup\mv@normal\amsa@group{msa}{m}{n}
      \define@mathgroup\mv@bold\amsa@group{msa}{m}{n}
      \edef\AMSa{\hexnumber\amsa@group}
      \makeatother
      \mathchardef\upi="0\UPM19
      \mathchardef\umu="0\UPM16
      \mathchardef\upartial="0\UPM40
      \mathchardef\leqslant="3\AMSa36
      \mathchardef\geqslant="3\AMSa3E

      \let\leq=\leqslant \let\le=\leqslant
      \let\geq=\geqslant 
    \fi
  \fi
\fi 

\ifnfsstwo
  \DeclareMathAlphabet{\mathbfit}{OT1}{cmr}{bx}{it}
  \SetMathAlphabet\mathbfit{bold}{OT1}{cmr}{bx}{it}
  \DeclareMathAlphabet{\mathbfss}{OT1}{cmss}{bx}{n}
  \SetMathAlphabet\mathbfss{bold}{OT1}{cmss}{bx}{n}
  \ifAMStwofonts
    \ifCUPmtlplainloaded \else
      \DeclareSymbolFont{UPM}{U}{eur}{m}{n}
      \SetSymbolFont{UPM}{bold}{U}{eur}{b}{n}
      \DeclareSymbolFont{AMSa}{U}{msa}{m}{n}
      \DeclareMathSymbol{\upi}{0}{UPM}{"19}
      \DeclareMathSymbol{\umu}{0}{UPM}{"16}
      \DeclareMathSymbol{\upartial}{0}{UPM}{"40}
      \DeclareMathSymbol{\leqslant}{3}{AMSa}{"36}
      \DeclareMathSymbol{\geqslant}{3}{AMSa}{"3E}

      \let\leq=\leqslant \let\le=\leqslant
      \let\geq=\geqslant 
    \fi
  \fi
\fi 

\ifCUPmtlplainloaded \else
  \ifAMStwofonts \else 
    \def\upi{\pi}
    \def\umu{\mu}
    \def\upartial{\partial}
  \fi
\fi

\title{Luminous early--type field galaxies at $z \sim 0.4$ \---\ I.
Observations and redshift catalogue of 581 galaxies}
\author[J. P. Willis et al.]
       {J.P.~Willis,$^{1}$\thanks{Present address: Departamento de Astromomia y Astrofisica, P. Universidad Catolica, Avenida Vicuna Mackenna 4860, Casilla 306, Santiago 22, Chile}
 P.C.~Hewett$^{1}$ and S.J.~Warren$^{2}$\\
       $^{1}$Institute of Astronomy, Madingley Road, Cambridge CB3
0HA\\ $^{2}$Blackett Laboratory, Imperial College of Science Technology and Medicine, Prince Consort Road, London SW7 2BZ}
\date{Accepted 2001 March 15.
      Received 2001 February 27;
      in original form 2000 October 21}

\pagerange{\pageref{firstpage}--\pageref{lastpage}}
\pubyear{2000}

\begin{document}

\maketitle

\label{firstpage}

\begin{abstract}
We have compiled a sample of $\sim 9600$ bright, $i\le 18.95$, red,
$b_j-r>2$, candidate galaxies in an area of 220 deg$^2$. These are
luminous, $L > L^*$, field early--type galaxies with redshifts $0.3 \la
z \la 0.6$.  We present a redshift catalogue of a sub--sample of 581
targets. The galaxies were selected according to their broadband
$b_jri$ colours from United Kingdom Schmidt Telescope plates, and have
a surface density on the sky of only $\sim 50\,$deg$^{-2}$. Such
luminous field galaxies are virtually absent from published redshift
surveys and the catalogue provides a large sample of the most luminous
normal galaxies, at cosmological distances. The statistical properties
of the galaxy spectra, including absorption line and emission line
measures, are presented and a composite spectrum constructed.  The nature
of the sample, combined with the relatively bright apparent magnitudes
make the galaxies suitable targets for several key investigations in
galaxy evolution and cosmology.

\end{abstract}


\section{Introduction}

Systematic galaxy redshift surveys to faint magnitude limits have
produced a wealth of information relating to the luminosity function
and evolution of the bulk of the galaxy population out to redshifts of
$z \sim 1$ (Ellis 1997).  Such studies have been complemented by
targeted investigations of rich clusters of galaxies out to a
comparable redshift (e.g. van Dokkum et al. 1998). Given the prevalence
of luminous ($L \ga L^*$) early--type galaxies within rich clusters it
has proved possible to obtain high--quality data for a large sample of
such galaxies in rich clusters.  By contrast, while the majority of
luminous early--type galaxies are located in field regions, where
``field'' is taken to indicate environments outside of rich galaxy
clusters, existing redshift surveys contain few examples of field
early--type galaxies with luminosities in excess of $L^*$ at redshifts
$0.2 \le z \le 0.8$.  Although such galaxies are intrinsically
luminous, and hence possess relatively bright apparent magnitudes,
their very low volume density leads to a surface density on the sky of
only $\sim 100\,$deg$^{-2}$. Deep redshift surveys, employing flux
limits at $R$--band or longer wavelengths, cover too small an area of
sky to include more than a few such objects. For example, only one of
the 591 galaxies in the CFRS survey with spectroscopic identifications
(Crampton et al.  1995) has a redshift $0.3 \le z \le 0.6$ and is
bright enough for inclusion in the sample presented here. Wide field
surveys, such as the 2dF Redshift Survey (Colless 1999), cover
sufficient area but the associated flux limits, often in a blue
passband, are bright. The integrated spectrum of an old passively
evolving stellar population exhibits a strong depression of the
continuum shortward of $4000\,$\AA \ resulting in a large
$k$--correction for blue passbands at redshifts $z \ga 0.25$, giving
observed colours $B-R > 2$. The brighter galaxies in the sample
presented here possess blue apparent magnitudes, $B \sim 21$, that are
still too faint by a magnitude or more for inclusion in the 2dF
Redshift Survey. Thus, notwithstanding the pioneering work of Hamilton
(1985) in identifying early--type galaxies at cosmological redshifts,
no large samples of the most luminous field early--type galaxies at
significant redshifts $z \sim 0.4$ exist.

The motivation for compiling a well--defined sample of luminous
early--type galaxies at redshift $z \sim 0.4$ is considerable, offering
the prospect of undertaking a number of projects not possible
hitherto.  Examples include: i) a survey for strong gravitational
lensing based on a study of the optimal deflector population (Hewett et
al. 2000), ii) investigating the Fundamental Plane for luminous galaxies
as a function of environment at a significant lookback time, iii)
establishing the evolution of the space density and luminosity function
of the most massive galaxies, and iv) quantifying the evolution of
galaxy clustering on large scales. Papers describing the
results of such investigations are in preparation and in this first
paper of the series we describe the procedures used to identify
luminous field galaxies with redshifts $0.3 \le z \le 0.6$ and present an
initial spectroscopic catalogue of 581 galaxies.

In Section 2 we describe the compilation of a sample of $\sim 9\,600$ bright,
red galaxies, $i\le 18.95$, $b_j-or>2$, over 220 deg$^2$, from scans of
United Kingdom Schmidt Telescope photographic plates. Multifibre
spectroscopic observations, using the 2dF instrument on the Anglo
Australian Telescope, of 581 of these candidate luminous early--type
galaxies are presented in Section 3. The section includes a discussion
of the reduction procedures employed to extract the galaxy spectra,
which are relatively faint by the standards of many observations made
with the instrument to date. The global properties of the galaxy
sample are investigated in Section 4. The redshift distribution and
number magnitude counts are presented and a composite spectrum
constructed. The absorption and emission line properties of the
galaxies are reviewed and the prevalence of [OII] 3727 emission
quantified. Further details of the absorption line strength
measurements are presented in the Appendix.  The paper concludes with
a summary of the properties of the galaxy sample in Section 5.

\section{Galaxy selection and photometric calibration}

This programme grew out of an earlier survey for high-redshift $z>2.2$
quasars (Warren et al. 1991), and the observational material and data
processing for the two surveys are very similar. Early--type galaxies
of redshift $z \sim 0.4$ and quasars of $z \sim4 $ have similar $B-R$,
$R-I$ colours, and the quasar samples were contaminated by galaxies.
The two populations may be distinguished on the basis of image profile
since the galaxies are extended and are marginally resolved on
the plates. Numerically the galaxies overwhelm the quasars, so for the
quasar survey it was essential to optimise the separation of resolved
and unresolved sources.

\subsection{Compilation of the $b_j$, $or$, $i$ multicolour dataset}

Relative to Galactic stars, early--type galaxies of redshift $z\sim0.4$
appear to be very red in $B-R$ but neutral in $R-I$, and may be
selected by identifying objects that lie to the red in $B-R$ of the
locus of early M--stars in a $BRI$ two--colour diagram. We have compiled
a sample of $9599$ bright, $i\le 18.95$, red galaxies, over 220.0 deg$^2$.
The basic photometric data used were the uncalibrated raw object
catalogues produced by Automated Plate Measuring (APM) machine scans of
photographic plates obtained at the United Kingdom Schmidt Telescope
(UKST). The survey covers seven UKST fields, listed in Table 1. A scan
of a single UKST field covers $33\,$deg$^2$ and the plates scanned are
listed in Table 2. The UKST passbands are defined by emulsion and
filter combinations. For the blue, red, and near--infrared the
combinations IIIa--J+GG395, IIIa--F+RG590, and IV--N+RG715 define the
$b_j$, $or$, and $i$ bands respectively (we use lower case for the UKST
photometric system). In field F833 we used the narrower $r$ band,
defined by IIIa--F+RG630, as there were no suitable $or$ plates. All
magnitudes are quoted in the natural system defined by the UKST
passbands, zero--pointed to the standard Johnson/Cousins system.

\begin{table}
\caption{UKST field centre coordinates.}
\label{tab_field}
\begin{tabular}{lccc}
Field & RA & Dec. & (B1950.0) \\
 & (hh:mm) & (dd:mm) & \\
SGP & 00:53 & -28:08 & \\
F297 & 01:44 & -40:00 & \\
F833 & 03:20 & +00:00 & \\
F855 & 10:40 & +00:00 & \\
F864 & 13:40 & +00:00 & \\
MT & 22:03 & -18:54 & \\
F345 & 22:06 & -40:00 & \\
\end{tabular}
\end{table}

The APM catalogues contain information on the brightness, shape, and
surface brightness profile of the objects detected, while the APM
instrumental magnitudes are related to standard magnitudes by a
calibration that is approximately linear, with a slope close to $-1.0$,
and a zero--point that is different for each plate. The steps involved
to produce clean calibrated multicolour data sets from the raw catalogues are
described in detail by Warren et al. (1991). For brevity, therefore,
only an outline of the data processing is provided here, except where
the procedure differed between the new galaxy survey and the earlier
quasar survey.

\begin{table}
\caption{UKST photographic plate numbers and galaxy magnitude limits for each
field.} 
\label{tab_field_limit}
\begin{tabular}{lcccccc}
Field & $b_j (1)$ & $b_j (2)$ & $r(1)$ & $r(2)$ & $i(1)$ & $i(2)$ \\
SGP  &  21.99 &  22.35 &  20.91 &  20.55 &  19.58 &  19.14 \\
SGP  &  J9766 &  J9771 & OR9563 & OR9595 & I12092 & I12198 \\
F297 &  22.17 &  21.49 &  20.37 &  20.62 &  19.30 &  19.33 \\   
F297 &  J3593 & J15720 & OR10353& OR10462& I10435 & I10441 \\
F833 &  22.29 &  21.99 &  20.66 &  20.36 &  20.09 &  19.79 \\
F833 &  J9752 & J14606 & R11438 & R11573 & I11437 & I16828 \\
F855 &  22.18 &  21.97 &  20.94 &  20.74 & 19.08  &  19.17 \\
F855 &  J9309 & J16076 & OR10854& OR14220&  I6910 & I10873 \\
F864 &  22.23 &  22.02 &  20.21 &  20.46 & 19.01  &  18.84 \\
F864 &  J9108 & J11701 & R6808  & OR13041& I10141 & I12562 \\
MT   &  22.23 &  22.24 &  20.34 &  20.48 &  18.89 &  19.28 \\ 
MT   & J12032 & J12129 & OR11946& OR12094&  I6389 &  I7058 \\ 
F345 &  22.52 &  22.40 &  20.66 &  20.64 &  19.18 &  19.90 \\   
F345 &  J2656 &  J3585 & OR12611& OR13849& I11430 & I12727 \\

\end{tabular}
\end{table}

The multicolour object catalogues are defined by objects that are
detected on both the red plates. The magnitudes were averaged after
transforming the instrumental magnitudes from one red plate to the
reference red plate. Then a search was made for these objects on the
plates of the other passbands, averaging the magnitude if detected on
both plates of a particular passband. If the object was undetected on
either plate the detection limit on the deeper plate was recorded. The
photometric errors, as a function of magnitude, were established
through an analysis of the scatter in the difference between magnitudes
of all the objects detected on the two plates of the same passband.
Fitting a polynomial function to this distribution provided a magnitude
error for each object, detected on each plate. The red plates are
sufficiently deep that all candidate early--type galaxies satisfying
the i--magnitude and colour cuts used to define the galaxy sample are
detected.

\begin{figure}
\psfig{figure=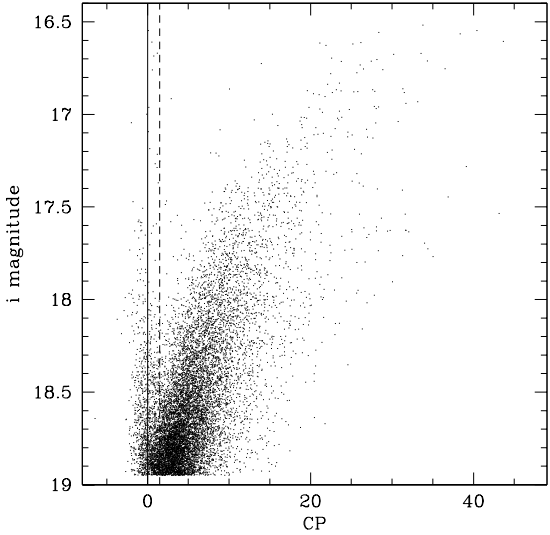,height=7.5cm,angle=0.0}
\caption{Classification parameter (CP) versus $i$ magnitude. The solid line
indicates the stellar locus defined to be ${\rm{CP}}=0$ and the dashed
line indicates the applied ${\rm{CP}}>1.5$ selection
threshold. Galaxian objects display a trend of greater CP values
at brighter apparent magnitudes.}
\label{cp_fig}
\end{figure}

The objects in the raw catalogues are classified as stellar or
non--stellar on the basis of a parameter computed from the
surface-brightness profile. This classification parameter (CP) is
rescaled by plotting the parameter against instrumental magnitude,
identifying the domain occupied by stars, zeroing to the peak, and
normalising by the standard deviation $\sigma$, such that for each
object on each plate CP now measures how many $\sigma$ the object lies
from the peak. The majority of the galaxies that are the target of this
work have half--light radii of order $1\arcsec$, and are difficult to
distinguish from stars on an individual plate. By averaging the values
of CP on all the plates on which an object was detected the separation
of stellar and non--stellar objects was greatly improved (Figure
\ref{cp_fig}).

A particular problem for the quasar survey was objects with incorrect
photometry, resulting in unusual (and therefore apparently
interesting) colours. The most common cause is two objects appearing
merged into a single object in one passband pairing with one of the
separate objects in another passband. To identify and eliminate such
cases we used two parameters; the positional offset between matched
objects on different plates, and the object ellipticity. By plotting
the distribution of these quantities against instrumental magnitude,
outliers were identified and eliminated. This procedure is appropriate
for stellar objects, but galaxies which have been matched correctly
might be eliminated by the ellipticity criterion. In fact this is less
of a problem for our targets because a) at these relatively faint
magnitudes the object ellipticity distribution is broader and the
cutoff therefore less strict, b) the galaxies have small angular sizes
due to their large distances and they appear close to circular due to
the seeing, and c) there are very few early--type galaxies with
intrinsic ellipticity greater than 0.6. To make this conclusion more
quantitative we reran the entire catalogue pipeline for one of our
fields, but without the ellipticity cut, and found that in the colour
region of interest the number of galaxies decreased by only $1\%$. We
conclude that the fraction of the total galaxy sample eliminated by
the ellipticity cut is negligible.

For the earlier quasar survey, that used 12 plates in each field, we
established that about $20\%$ of real sources were removed by the
procedures designed to eliminate objects with incorrect photometry
(e.g. object mergers, emulsion flaws and satellite trails). Assuming
that the incidence of objects with incorrect photometry is consistent
across all plates then for the current galaxy survey, which uses only
six plates per field, we calculate that $\sim 10\%$ of real sources
are lost in this way. The effective area of the survey is therefore
reduced from 220.0 deg$^2$ to 198.0 deg$^2$.

At this point the catalogue comprises all cleanly matching objects
detected on both the red plates, with APM magnitudes, or limits, in the
three bands, as well as photometric errors and an average value of CP.
The object x--y positions in the APM catalogues are transformed to
celestial J2000.0 coordinates via a plate--solution based on the right
ascension, declination and proper motions of the $\sim 600$ TYCHO--2
catalogue stars (H{\o}g et al. 2000) present on each plate. The
root--mean--square positional differences between the catalogued
TYCHO--2 star postions and the APM--positions is $\simeq
0.25\,$arcsec. Systematic errors in positions, relative to the frame
defined by the TYCHO--2 stars, are $\la 0.5\,$arcsec across a full UKST
field.


\subsection{Photometric calibration}

Photometric calibration of galaxies on photographic plates is
complicated.  This is because APM instrumental magnitudes are
essentially isophotal magnitudes, but calibrated aperture or total
magnitudes are required. The difference between APM isophotal magnitude
and aperture magnitude depends on the surface brightness profile of the
galaxy, and this inevitably introduces scatter into the calibration, in
addition to the photometric errors. In other words the $\chi^2$ for the
fit of the calibration--curve will be large. The problem is severe when
using a calibration sequence containing all galaxy types, as for
example in the APM galaxy survey (Maddox, Efstathiou \& Sutherland
1990). The problem is much less severe when calibrating a homogeneous
population using CCD measurements of galaxies drawn from the
population, as is the case here. Considering this question in more
detail, for the population of giant early--type galaxies there is a
strong correlation between physical size (e.g. half--light radius) and
luminosity.  Because the redshift distribution of the sample is quite
narrow (Figure 17), this results in a strong (anti--)correlation
between size and apparent magnitude. Therefore the magnitude correction
from isophotal to aperture magnitude is also strongly correlated with
apparent magnitude. Consequently the increased scatter in the
calibration curve, due to the spread in magnitude correction at any
apparent magnitude, is small. This conclusion is borne out by the
measured $\chi^2$ for our calibration curves, and we conclude that the
aperture magnitudes are quite reliable.

We obtained CCD images of galaxies in five fields in 1993 November at
the Danish 1.5m and ESO 3.6m telescopes on La Silla. Integration times
of $1500\,$s, $300\,$s and $240\,$s were used for the $B$, $R$, and
$I$ bands respectively for the 1.5m, and $300\,$s, $180\,$s and
$180\,$s for the 3.6m. CCD images in fields F855 and F864 were obtained
in 1997 April with the 2.5m Isaac Newton Telescope on La Palma.
Integration times were $1400\,$s, $400\,$s and $400\,$s for the $B$,
$R$, and $I$ bands respectively.  An average of 23 galaxies (range 12
to 46) were observed in each field. Standard procedures for data
reduction were followed in processing these data. Magnitudes were
measured using an $8\arcsec$ diameter aperture. To convert the Johnson/Cousins
magnitudes to UKST magnitudes requires colour equations suitable for
the objects in question. For this purpose we computed synthetic colours
of early--type galaxies in the redshift range $0.3 \le z \le 0.6$,
obtaining the following linear fits for the colour terms:

\[ b_j-B=1.233-0.593(B-R) \]
\[ or-R=-0.307+0.337(R-I) \]
\[ r-R=-0.212+0.056(R-I) \]
\[ i-I=0.0 \]

The galaxy CCD magnitudes were transformed to the UKST natural system,
and a least--squares linear calibration solution derived. Typical
errors for the standards were 0.08, 0.02, 0.03 mag in $B$, $R$, $I$
respectively, and 0.13, 0.06, 0.12 for the photographic magnitudes. A
weighted linear solution was computed, assuming errors dominated by the
photographic magnitudes, providing a satisfactory fit over the limited
magnitude range of the sample. A correction for Galactic reddening of
$E(B-V)=0.05\,$mag was made for F833 which has a Galactic latitude $b
\sim -45^{\circ}$. Using this calibration the number of galaxies meeting
the colour selection criteria showed rather large variations from field
to field. Therefore, for fields with few galaxy standards we permitted
small adjustments to the calibration zero--points, of $\le 0.1\,$mag.
Adjustments were determined by a careful comparison of the mean galaxy
colours, and the surface density of the colour--selected galaxy
samples. The accuracy of the adopted zero--point in each field is thus
$\simeq \pm 0.05\,$mag.

\subsection[]{Sample selection}
\label{sec_sample_select}

\begin{figure}
\psfig{figure=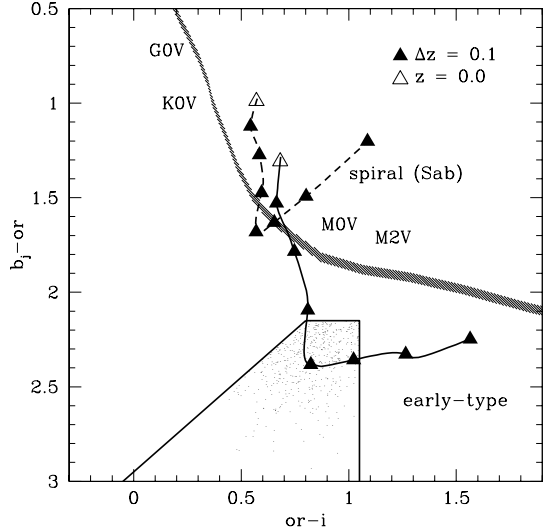,height=7.5cm,angle=0.0}
\caption{$b_jori$ two--colour diagram indicating the relative position
of loci representing evolving early--type galaxies (solid curve) and
spiral galaxies (dashed curve) versus redshift (see text for
additional details), and the stellar main sequence (shaded
region). The boundaries employed to select early--type galaxy
candidates are shown by the solid polygon at the lower--middle of the
plot. Individual candidate galaxy colours are indicated by small dots.}
\label{ms_colour}
\end{figure}

Early--type galaxies are selected according to the CP value and
location in $b_jori$ colour space. Colour--selection on the $b_j-or$
versus $or-i$ colour plane is a highly effective method of isolating
early--type galaxies from both stars and later galaxy types. The locus
occupied by early--type galaxies is largely determined by the effect
of the 4000{\AA} break moving through the $b_j$ passband with
increasing redshift. Figure \ref{ms_colour} shows the locus defined by
an evolving early--type galaxy spectral energy distribution (Pozzetti
et al. 1996).  The locus defined by an evolving Sab spiral galaxy
spectral energy distribution (Pozzetti et al. 1996) is shown for
comparison. Each locus was created by specifying a redshift of galaxy
formation $z_{form}=4.5$ and a cosmological model described by the
parameters $\Omega_M=0.3$, $\Omega_{\Lambda}=0.7$ and $H_0 = 70$
kms$^{-1}$ Mpc$^{-1}$. Dispersion about the locus arises from random
photometric errors and intrinsic colour variation within the galaxy
population.  Defining the multi--colour parameter space listed in
Table \ref{tab_photo_select} isolates early--type galaxies in the
redshift interval $0.3 \la z \la 0.55$. A random error $\ga 5 \sigma$
in the measurement of early--type spiral colours is necessary to
scatter such galaxies into the region defined by the colour--selection
criteria.  The locus of stellar colours on the $b_j-or$ versus $or-i$
plane is defined by the main sequence and falls outside the defined
colour selection region.  However, dispersion about the locus could
lead to contamination of the early--type galaxy sample by M--type
stars. The stellar main sequence is sufficiently removed from the
nearest colour--selection boundary that contamination of the
early--type galaxy sample is likely to be low. Furthermore, the
requirement that candidate galaxies possess a morphological
classification CP $>$ 1.5 further reduces the potential stellar
contamination. The effectiveness of the morphological selection is
addressed in Section 3.4, where the results of spectroscopic
observations of a sample of objects satisfying the colour--selection
criteria but possessing morphological classifications CP $<$ 1.5 are
presented.

\begin{table}
\caption{Photometric selection criteria.}
\label{tab_photo_select}
\begin{tabular}{ccccc}
16.5 & $<$ & $m_i$ & $<$ & 18.95 \\
2.15 & $<$ & $ b_j - or $ & $<$ & 3.00 \\
& & $ or - i $ & $<$ & 1.05 \\
2.95 & $<$ & $ b_j - i $ & & \\
\end{tabular}
\end{table}

\begin{figure}
\psfig{figure=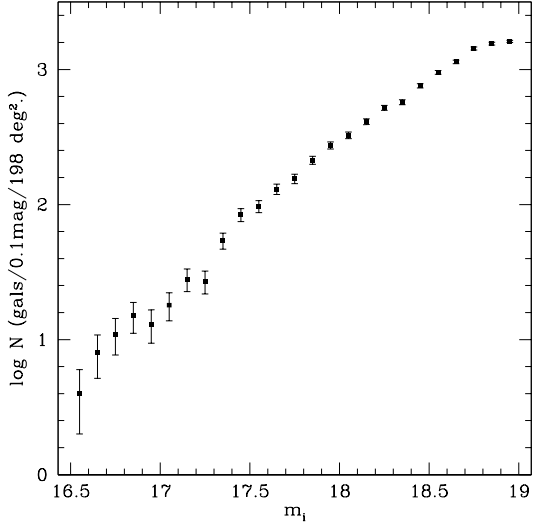,height=7.5cm,angle=0.0}
\caption{Number--apparent magnitude distribution for
early--type galaxy candidates in the seven UKST fields. Poisson
error bars are shown.}
\label{survey_nm}
\end{figure}

The selection procedure as described above was applied to the seven UKST fields,
identifying 9599 early--type galaxy candidates from an effective area
of $198.0\,$deg$^2$. The surface density of objects at the faint
magnitude limit ($m_i \le 18.95$) is $\sim 50\,$deg$^{-2}$. Figure
\ref{survey_nm} shows the $i$--band number--apparent magnitude relation
for the sample.

\section[]{Spectroscopy}

\subsection[]{Observations}

Spectroscopic observations were performed using the Two--degree Field
(2dF) multi--fibre spectrograph at the Anglo--Australian Telescope
(AAT). With a surface density of $\sim 50\,$deg$^{-2}$ up to 150
candidate galaxies can be observed simultaneously with 2dF. In addition
to observations of early--type galaxy candidates, a sample of 33
stellar candidates (i.e. objects satisfying the colour criteria but
with ${\rm{CP}}<1.5$) were included in the 1998 September 16--17 2dF
observations in order to determine directly the level of incompleteness
in the early--type galaxy sample.

Target 2dF fields were configured using the observatory supplied
software package {\tt configure} (Bailey and Glazebrook 1999) which
automatically allocates fibre positions to guide star, early--type
galaxy and stellar targets. Sky fibres were allocated according to the
automated {\tt skygrid} procedure within {\tt configure}. Sky fibres
were allocated in excess of the conventional rate (Wyse and Gilmore
1992) as previous experience with 2dF observations indicated that
systematic instrumental effects (spectrograph focus, scattered light)
rather than photon noise, limit the accuracy of sky--subtraction.

\begin{table}
\caption{Spectroscopic observations.}
\label{tab_spectro_obs}
\begin{tabular}{llccccc}
Date & Field & Exposure time (s) & $N_{spectra}$ \\
98/09/16--17 & F297, MT & 12,600 & 485 \\
98/05/03 & F864 & 6,300 & 18 \\
97/07/31 & F864 & 7,200 & 78 \\
\end{tabular}
\end{table}

The spectrograph configuration was essentially identical for all the
observing runs, employing the AAO 600V gratings to give spectra
covering the wavelength range $4600-6800$\AA \ with a resolution of
$\sim 4$\AA.  Observations of each target field were made up of
individual exposures of $1800-2100\,$s. Tungsten--lamp flat--field
exposures and several CuHe+CuAr arc calibration exposures were taken
for each target field.  Offset sky exposures were also obtained but
were not in fact used in the reductions. A summary of the spectroscopic
observations are given in Table \ref{tab_spectro_obs}. The spectra
presented in Section 4 were obtained from observations made over the
two--nights of 1998 September 16--17 (Table 5).  Atmospheric
transparency was good throughout and the seeing varied between
$1\farcs5$ and $2\farcs0$.  Aside from some contaminating light
from an uncovered Light Emitting Diode affecting spectrograph B in
exposures obtained for one field (see next sub--Section) no significant
problems occurred with the telescope or the 2dF instrument. The
discussion of the reductions and the analysis of the spectroscopic
properties of the galaxy sample presented in subsequent sections is
based on the data obtained during this run.

\begin{table*}
\caption{2dF Field coordinates and target classifications for 1998/09/16--17 observations.}
\label{tab_obs_field}
\begin{tabular}{cccccccc}
Field & \multicolumn{2}{c}{Centre (J2000)} & \multicolumn{5}{c}{Fibre allocations} \\
 & RA (hh:mm:ss.s) & dec. (dd:mm:ss) & Early--type & Stellar & QSO &
 Sky & Parked \\
MTF 2 & 22:11:58.3 & -20:33:07 & 141 & 8 & 74 & 102 & 75 \\ 
MTF 5 & 21:56:27.2 & -20:47:20 & 114 & 17 & 75 & 108 & 87 \\ 
F297 1 & 01:45:59.2 & -39:51:34 & 135 & 4 & 0 & 200 & 61 \\ 
F297 2 & 01:35:15.6 & -40:05:10 & 135 & 4 & 0 & 208 & 43 \\
\end{tabular}
\end{table*}

Observations made prior to the 1998 September run were obtained with
the 2dF instrument still not fully commissioned and the weather in all
cases produced periods of variable transparency. As a result only some
of the exposures obtained were of use and spectra were obtained for
only a subset of the targets. However, the spectra of a significant
number of galaxies observed were of sufficient quality to yield
reliable redshifts. These spectra were not included in the sample
employed for statistical analysis but the positions, magnitudes and
redshifts of the galaxies are included in Table \ref{spec_cat2}.

\subsection[]{2dF data reduction}

Spectra were reduced using the AAO--supplied {\tt 2dFdr} package and
standard {\tt IRAF}\footnote{IRAF is distributed by the National
Optical Astronomy Observatories, which are operated by the Association
of Universities for Research in Astronomy, Inc., under cooperative
agreement with the National Science Foundation.} routines. All spectra
were inspected visually prior to applying reduction
techniques. Contaminating light from an uncovered Light Emitting Diode
(LED) affecting spectrograph B (CCD2) was noted in exposures obtained for
one field (MTF5). The contaminating signal includes a broad ($\sim
1000${\AA}) emission feature located at $\sim 6000${\AA}. The effect
of this contamination on the reduction methods is discussed in
subsequent sections. Spectra were de--biased, scattered light
subtracted, extracted and wavelength calibrated using the {\tt 2dFdr}
package.  Correction of fibres to a uniform transmission level and
sky--subtraction were performed using specially written routines in
the {\tt IRAF} environment.

\subsection[]{Sky--subtraction}

The signal received through a 2dF fibre from a galaxy target is a
factor $\sim 3$ fainter than the $R$--band night--sky. Accurate
determination of relative fibre transmission levels and subtraction of
the night--sky contribution are essential to produce high quality
object spectra. The sky conditions were stable over the observations of
individual fields. As the spectrograph exhibited a high degree of
stability, with wavelength shifts over the observation of individual
fields corresponding to considerably less than one CCD--pixel, the
individual exposures of each target field were combined prior to
correction for fibre transmission and sky--subtraction. Spectra were
co--added within {\tt 2dFdr} applying a $5 \sigma$--clipping rejection
criterion to remove cosmic ray events.

Relative fibre transmission levels were estimated by combining the
emission--line signal in prominent night--sky emission--lines located
at 5577{\AA}, 5890{\AA} and 6300{\AA} for each object+sky and sky fibre
(Lissandrini, Cristiani and La Franca 1994).  The ``continuum''
contribution from the sky and object in each fibre spectrum was removed
by fitting and subtracting a 21st order cubic spline function. The
narrow absorption features in early--type galaxy spectra are not strong
enough to bias significantly the estimate of the night--sky emission
signal. Use of the night--sky emission line strength method is
relatively unaffected by uncertainties in the scattered light
correction applied to the frames and offers a significant advantage
over the use of offset sky exposures to estimate the relative fibre
transmissions. Employing a relatively wide ($\sim 50\,$\AA) window to
calculate the signal associated from night--sky emission lines
minimises systematic variation of the line--to--continuum ratio caused
by varying spectrograph focus. The standard {\tt 2dFdr} method (option
{\tt SKYLINE}) was found to be susceptible to this effect. Individual
fibre spectra were then corrected to a uniform transmission level.

Variations in spectrograph focus along the slit ($\sim20${\%} from the
centre to the edge of particular fields) and hence with fibre number,
limit the effectiveness of a sky--subtraction method based upon a
single template sky spectrum. Therefore, a sky--spectrum template was
formed from $\sim 20$ sky--fibre spectra with the best focus on each
combined exposure.  The resulting template sky--spectrum was then
convolved with a Gaussian profile of varying width, chosen to match the
observed Full--Width Half--Maximum (FWHM) of prominent night--sky
emission--lines in each individual object+sky-- and sky--fibre
spectra.

\begin{figure}
\psfig{figure=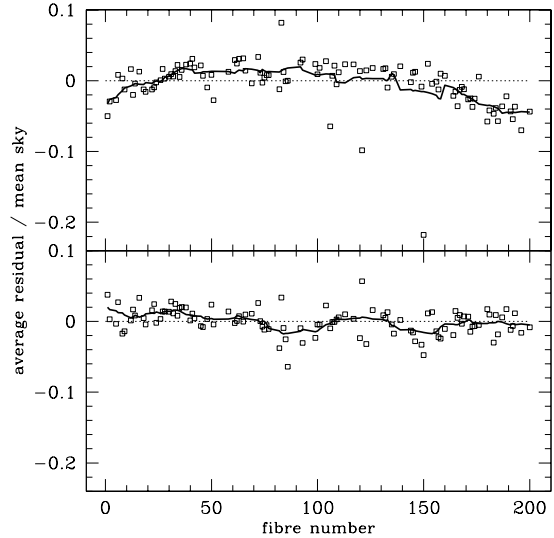,width=7.5cm,angle=0.0}
\caption{Top panel: average residual for {\tt 2dFdr} sky--subtracted
sky--fibres in Field F297 2 CCD2 (observed on 1998/09/17). The solid
line is obtained by median filtering the data over an 11--fibre window.
The dotted line indicates zero residual. Bottom panel: average residual
per pixel obtained using the modified sky--subtraction procedure
described in the text.  The rms deviation about the zero level is
equivalent to a $\pm2{\%}$ sky--subtraction uncertainty.}
\label{sky_residual}
\end{figure}

\begin{figure}
\psfig{figure=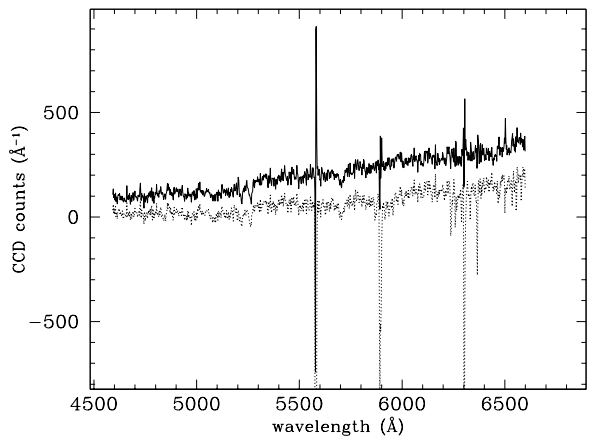,width=8.0cm,angle=270.0}
\caption{Example of an early--type galaxy spectrum (solid line) reduced
using the modified sky--subtraction procedure compared to the spectrum
of the same object reduced using the standard {\tt 2dFdr}
sky--subtraction method (dotted line).}
\label{recover_spec}
\end{figure}

The effectiveness of the fibre--transmission and sky--subtraction
procedures compared to the standard {\tt 2dFdr} reduction is
illustrated in Figure \ref{sky_residual} and Figure
\ref{recover_spec}.  The systematic trends in residual zero--point
level following sky--subtraction are much reduced and individual
examples of poor sky--subtraction can show dramatic improvements. The
accuracy of sky--subtraction, calculated from the dispersion in the
zero--point of sky--subtracted sky--fibres relative to the mean in a
given frame, is $\pm 2{\%}$. The dispersion is significantly larger than
the theoretical Poisson limit, $< 1\%$, but represents a significant
improvement over that achieved, $\sim 3-5\%$, through employing the default reduction
procedures in {\tt 2dFdr}.

\subsection[]{Spectral classification}
\label{subsec_classify}

At this point the 2dF spectra consist of wavelength calibrated
one--dimensional sky--subtracted spectra, each of which has an
associated classification (``galaxy'', ``stellar'', ``sky'') derived
from the input catalogue.

\subsubsection[]{Stellar candidates}

Comparison of each of the spectra of the 33 stellar targets to
reference spectra of cool main--sequence stars (Turnshek et al. 1985) produced
unambiguous identifications as late K and early M stars. Individual
spectra showed evidence for some, or all, of broad molecular absorption
bands, such as TiO, and narrow absorption lines (MgH4780, Mgb, NaD and
CaH) superimposed upon a characteristic continuum shape. The sample of
stellar candidates does not contain any galaxies, showing that the
incompleteness in the early--type galaxy sample due to
misclassification of galaxies as stars is extremely small.

\begin{figure}

\psfig{figure=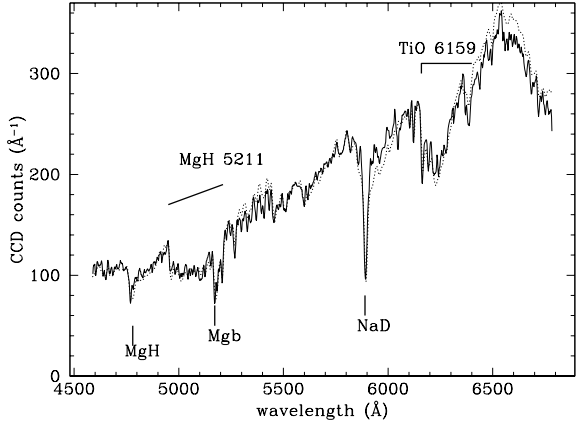,width=8.0cm,angle=270.0}
\caption{Mean stellar spectrum (solid line) generated from 33
individual response function corrected spectra (see Section 3.6)
compared to a template M2V spectrum from the atlas of Pickles (1998)
(dotted line). Strong absorption features are indicated.}
\label{mstar_temp}

\end{figure}

A composite spectrum was generated from the stellar spectra.  Spectra
were combined using inverse--variance weighting and 3$\sigma$--clipping
rejection criteria to remove individual pixels affected by residual
cosmic rays and night--sky emission lines. Strong residual signatures
associated with the very strong night--sky emission--lines at
5577{\AA}, 5890{\AA}, 6300{\AA} and 6360{\AA} were replaced by a local
median--filtered continuum estimate. Figure \ref{mstar_temp}
indicates that the composite spectrum corrected for the instrumental
response (Section \ref{subsec_response}) closely resembles a star of
type M2V (Turnshek et al. 1985; Pickles 1998).

\subsubsection[]{Galaxy candidates}

Classification of the spectra of candidate galaxies was accomplished
via cross--correlation with representative early--type galaxy and
stellar templates. A high signal--to--noise ratio (S/N) early--type
galaxy spectral energy distribution (SED) derived from observations of
local early--type galaxies (Kinney et al. 1996) and the composite stellar
spectrum derived above were employed as templates.  The residual
signature of strong night--sky emission--lines in the object spectra
were excluded from the cross-correlation analysis to avoid introducing
spurious signals. Individual spectra were cross--correlated with each
template using the {\tt IRAF} routine {\tt xcsao} (Tonry and Davis 1979). This
routine effectively removes continuum shape information extending over
large wavelength intervals and calculates the cross--correlation
signal between continuum discontinuities and absorption and emission
features in the spectra. The cross--correlation can be performed over
a range of radial velocities in order to ascertain the radial velocity
(redshift) for the template that maximises the cross--correlation
signal.  Cross--correlation with the stellar template was performed
over the radial velocity interval $\pm500\, \rm{km}\,\rm{s}^{-1}$.
Cross--correlation with the galaxy template was performed over the
radial velocity interval $0-200000\, \rm{km}\,\rm{s}^{-1}$ to allow
for early--type galaxy redshifts $z \le 0.67$ (exceeding the maximum
anticipated redshift).

\begin{figure}
\psfig{figure=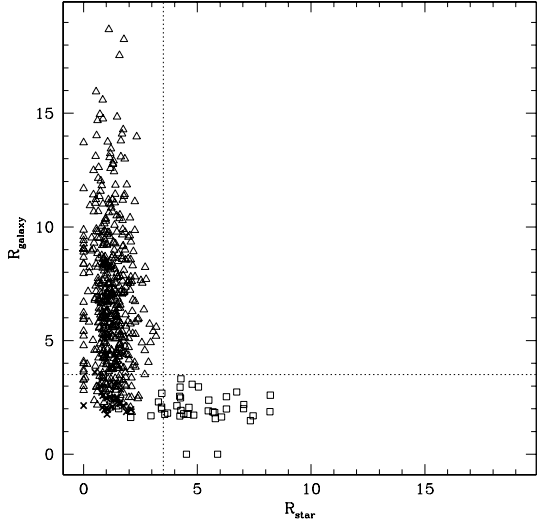,width=7.5cm}
\caption{R--values for the 525 2dF spectra derived from
cross--correlation with stellar (X--axis) and galaxy (Y--axis)
templates using {\tt xcsao}. Adopted spectroscopic classifications are
indicated for early--type galaxies (open triangles), stars (open
squares) and unclassified spectra (crosses).}
\label{rr_class}
\end{figure}

\begin{figure}
\psfig{figure=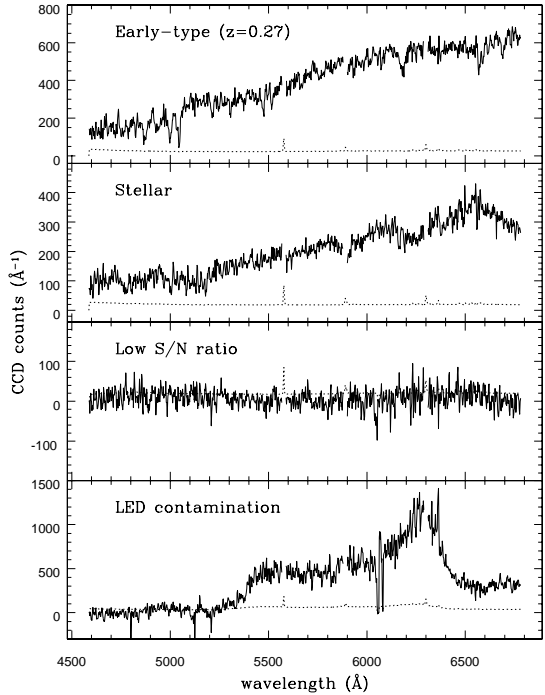,width=7.5cm}
\caption{Example spectra drawn from each spectral class. The dotted
line indicates the 1$\sigma$ noise level in each case.}
\label{class_examples_port}
\end{figure}

Figure \ref{rr_class} displays the resulting R--value\footnote{The
R--value describes the ratio of the cross--correlation peak height to
(twice) the antisymmetric noise component and offers an estimate of
the significance of the cross-correlation peak.} for each template
applied to each spectrum from the 1998/09/16-17 observations. Stellar
spectra lie below $\rm{R_{galaxy}}=3.5$ and, for $\rm{R_{star}}>3.5$,
are clearly separated from candidate early--type galaxy spectra.  This
leaves a potential zone of confusion containing 126 spectra with
$\rm{R_{star}} \la 3.5$ and $\rm{R_{galaxy}} \la 3.5$. Although these
spectra generally display low--S/N, spectral information can still be
used to discriminate between galaxies and stars. Candidate early--type
galaxy spectra were classified visually by identifying CaII(H+K),
G--band, Mgb and Balmer absorption observed at $0.25 \la z \la 0.63$
(86 objects). Spectra displaying absorption and continuum properties
consistent with the mean stellar template were classified as stars (8
objects). Spectra that could not be classified (32 objects) fell into
two groups:  low--S/N spectra displaying no identifiable spectral
signature and spectra displaying significant contamination from
extraneous LED light or instrumental defects. Table \ref{tab_spec}
summarises the number of objects identified in each class for the
1998/09/16--17 2dF observations.  Examples of classifications from
each category are shown in Figure \ref{class_examples_port}. The
stellar contamination of the early--type galaxy sample is $\simeq
2${\%}.

\begin{table}
\caption{Classification of candidate early--type galaxy spectra in the
1998/09/16--17 sample.}
\label{tab_spec}
\begin{tabular}{lc}
Early--type galaxy & 485 \\
Stellar & 8 \\
Unclassified (low S/N) & 24 \\ 
Unclassified (LED/defect) & 8 \\
Total & 525 \\
\end{tabular}
\end{table}

The sample of early--type galaxy spectra was sub--divided to account
for variations in data quality. Sample A (433 objects) contains all
spectra with reliable continuum shapes. This excludes spectra from MT
field 5 observed using CCD2 which display evidence of contamination by
LED light. Sample A is used to determine continuum--dependent
properties, i.e. the composite spectrum and 4000{\AA} break
indices. Sample B (485 objects) contains all spectra for which reliable
redshifts could be determined.

\subsection[]{Redshift determination}
\label{sec_red_determ}

The position of the cross--correlation peak indicates the galaxy
redshift. However, for 19 spectra, with R$\la$6, the cross-correlation
method proved unsatisfactory, due either to weak absorption features or
residuals due to bad pixels, cosmic--rays or night--sky residuals at
the wavelengths of major absorption features (e.g. CaII (H+K), G--band
and Mgb). These objects were assigned redshifts interactively using the
4000{\AA} break and less prominent absorption features (e.g. H$\delta$,
H$\beta$ and Fe lines).

As a check of the redshift assignments, each spectrum was
de--redshifted to the rest--frame and the wavelength interval
3500--4500{\AA} plotted. This permitted a rapid visual check of the
redshift of each spectrum by noting the position of the strong CaII H+K
absorption plus the 4000{\AA} break feature.

Errors in the cross--correlation velocity returned by the {\tt IRAF}
routine {\tt xcsao} are computed based on the fitted peak height and
antisymmetric noise associated with the identified cross--correlation
peak (Tonry and Davis 1979; Heavens 1993). As expected with moderate
to low S/N data, the error in the returned velocity estimate is
a strong function of galaxy spectral S/N (Figure
\ref{gal_z_err}). The base level of error is $\sim 35$ kms$^{-1}$ yet
the sample displays a ``tail'' of velocity errors $\la 200$
kms$^{-1}$ at S/N $\la 10$.

\begin{figure}
\psfig{figure=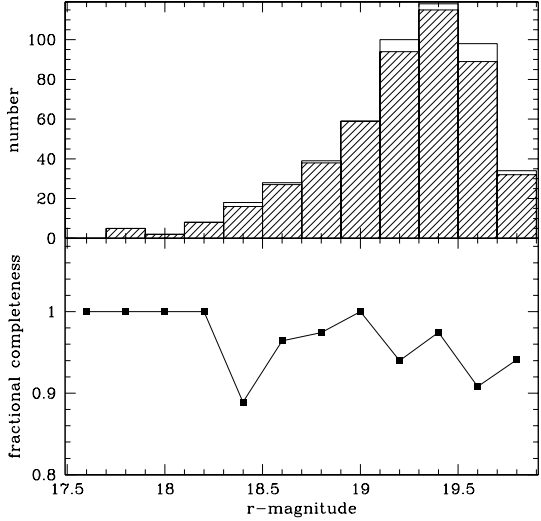,width=7.5cm,angle=0.0}
\caption{Top panel: Number--magnitude histogram. The open histogram
indicates the number of objects in each apparent magnitude bin. The
shaded histogram indicates the number of objects allocated a redshift.
Bottom panel: Fractional redshift completeness versus apparent
$r$--magnitude.}
\label{z_completeness}
\end{figure}

\begin{figure}
\psfig{figure=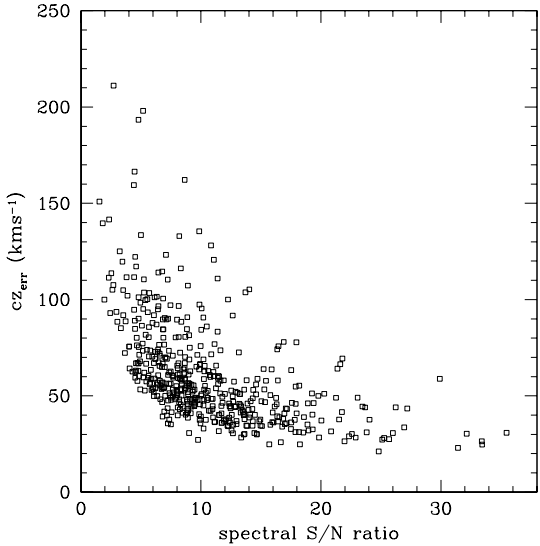,width=7.5cm,angle=0.0}
\caption{Galaxy redshift velocity error as a function of spectral S/N.}
\label{gal_z_err}
\end{figure}

Some variation in redshift completeness as a function of apparent
magnitude is expected.  However, there is no evidence for any
differences in the fractional redshift completeness rate for galaxies
observed on different nights or in the two different fields. Therefore,
a redshift incompleteness correction as a function of apparent
magnitude only is derived. The number of the 525 early--type galaxy
targets identified as stars (8) constitutes a very small percentage,
$\simeq 2\%$, of the total and the majority of the 32 objects that
remain unidentified are almost certainly early--type galaxies. Assuming
this to be the case, Figure \ref{z_completeness} displays redshift
completeness as a function of apparent $r$--magnitude for the sample of
517 objects comprising the early--type galaxies and unidentified
spectra. The fractional completeness is extremely high ($0.95$) and
shows only a weak dependence on apparent magnitude. In fact, the low
S/N and LED--contaminated spectra making up the unidentified objects
result primarily from factors such as low fibre thruput, astrometric
errors, and fibre placement errors, unrelated to the target
magnitudes.

\subsection[]{Flux calibration}
\label{subsec_response}

The homogeneous nature of the target galaxy population enables a highly
effective ``bootstrap'' procedure to be employed to determine a
relative flux scale. The procedure is based on the comparison of each
object classified as an early--type galaxy to a reference early--type
galaxy template (Kinney et al. 1996).  The template spectrum was
redshifted and resampled to provide a match to the observed--frame
spectrum of each early--type galaxy. The template was then
appropriately normalized and divided into the galaxy spectrum. The
overall similarity of the resulting $\sim 60$ individual response
functions per field+CCD confirmed the homogeneous nature of the galaxy
population. The median response per pixel (i.e. wavelength) was
calculated and a third--order cubic spline function fitted to produce a
response function for each field+CCD combination.

The response function for each field+CCD was applied to all spectra
observed with the respective field+CCD combination. The form of the
response function accords well with expectations based on the
properties of the atmospheric transmission, spectrograph grating
efficiency and CCD quantum efficiency. However, an independent test to
show that the reference template is indeed appropriate is desirable.
The inclusion of the 33 stellar objects with colours consistent with
M--stars scattered from the stellar locus (Figure \ref{ms_colour})
allows such a test to be performed. A composite stellar spectrum was
generated from the 33 individual response function--corrected spectra.
Figure \ref{mstar_temp} shows the ``flux--corrected'' mean stellar
spectrum together with a reference M2V spectrum. The spectral type has
been chosen to produce the best match to the absorption features, not
to optimise the match of the continua. The agreement in the shape of
the continua over intermediate and large scales indicates that the
bootstrap procedure based on the use of a reference early--type galaxy
template produces a relative flux scale accurate to $\sim 10\,\%$.
Such an accuracy is typical for observations using narrow slits over
wavelength intervals of $\sim 2000\,$\AA \ and can be achieved for the
2dF observations because of the extremely homogeneous nature of the
galaxy target spectra and their similarity to the $z \sim 0$
early--type galaxy template.

An absolute flux zero--point for each spectrum can be derived using the
apparent magnitudes after allowing for the fraction of light from the
galaxy, modeled as a de Vaucouleurs (1948) surface brightness
distribution with half--light radius $1 \scnp$, entering the $2 \scnp$
diameter fibres in the seeing conditions appropriate to the 2dF
observations. The conversion factor between the UKST broad--band
$i$--magnitudes and the flux present in the spectral interval
4800--6600{\AA} \ was determined from a suitably redshifted version of
the Kinney early--type galaxy template, together with the UKST
$i$--band filter and photographic emulsion response functions.

\section[]{Statistical properties of the sample}

\subsection[]{The composite early--type galaxy spectrum}
\label{sec_comp_spec}

The early--type galaxy sample is sufficiently homogeneous that the
properties of the composite spectrum are potentially of considerable
interest.  To generate a composite spectrum, all early--type galaxy
spectra in Sample A (Section \ref{subsec_classify}) were de--redshifted
and resampled, using $1.5\,$\AA \ pixels, onto the rest--frame
wavelength interval 2900--5405{\AA}. An associated quality array
specified the location of bad CCD pixels, missing data and residual
signatures associated with prominent night--sky emission features.
Wavelengths so flagged in the quality array were not included in the
generation of the composite.

Prior to combination, each spectrum was normalised to a uniform mean
count value of unity over the rest--frame wavelength interval
3670--4180{\AA}, the interval common to all spectra.  An initial
composite spectrum and associated 1$\sigma$ noise estimate was
calculated according to:

\begin{equation} 
\overline{R}_j = \frac{1}{n_{spec}} \sum_{i=1}^{n_i} k_i R_{ij},
\label{eqn_one}
\end{equation}

\begin{equation} 
\overline{N}_j = \left [ \frac{1}{n_{spec}} \sum_{i=1}^{n_i} {( k_i N_{ij} )}^2
\right ]^{\frac{1}{2}},
\label{eqn_two}
\end{equation}

{\noindent}where $\overline{R}_j$ is the composite spectrum over
pixels $j=1, n_{spec}$. $R_{ij}$ is the jth pixel of the ith
spectrum. $k_i$ is the normalisation constant applied to each
spectrum. $\overline{N}_j$ and $N_{ij}$ are the corresponding
1$\sigma$ noise values per pixel.

The initial estimate of the composite spectrum was used as a template
to identify any remaining bad pixels not included in the quality
arrays (e.g. poorly removed cosmic rays). Each spectrum was divided by
a suitably scaled version of the composite spectrum to generate an
exact transformation between the spectra. The transformation array was
median--filtered with a 31--pixel window ($\sim 70${\AA}) to remove
positive and negative deviations on wavelength scales $\la 30${\AA}
yet preserve large--scale continuum variations. The composite was then
scaled using the median--filtered transformation array and subtracted
from the individual spectrum to produce a residual spectrum. Pixels
deviating by $>$5$\sigma$ from zero in each residual spectrum were
added to the quality array. Very few pixels ($\la 5$) were rejected
from individual spectra via this step. To investigate the homogeneity
of the sample quantitatively, the normalised $\chi^{2}$ statistic was
calculated for each individual spectrum and the scaled composite over
the full rest--frame wavelength interval:

\begin{equation} 
\chi^{2}_{i} = \frac{1}{n_{pix}}
               \sum_{j=1}^{n_{pix}}
               \frac{(R_{ij}-\overline{R}_j/k_{i})^2}
               {N_{ij}^2}. 
\label{eqn_three}
\end{equation}

\begin{figure}
\psfig{figure=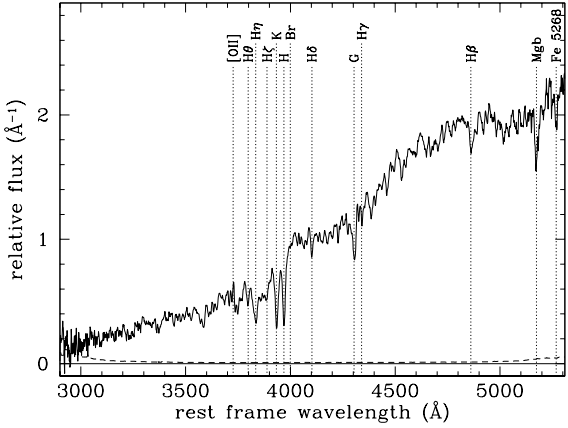,width=8.0cm,angle=270.0}
\caption{The composite spectrum
resulting from four iterations of $\chi^2$ rejection algorithm. The
dashed line indicates the associated $1\sigma$ noise array. The solid
horizontal line indicates the zero relative flux level. Prominent
rest--frame absorption features are indicated by vertical dotted lines.}
\label{mspec}
\end{figure}

\begin{figure}
\psfig{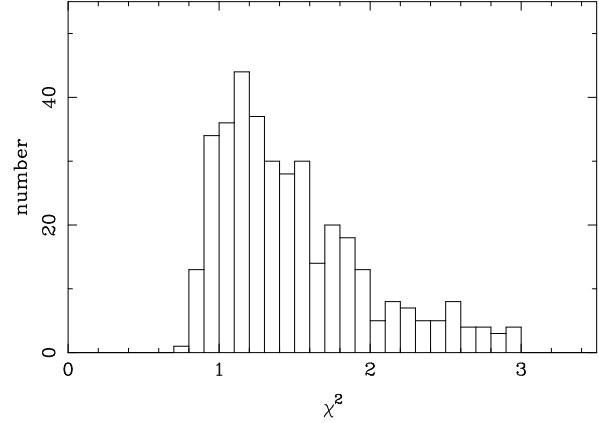}
\caption{Histogram of $\chi^2$ values
after four iterations of the rejection algorithm.}
\label{chi_four}
\end{figure}

Given the goal of obtaining a composite spectrum representative of the
homogeneous element of the sample, objects with $\chi^2 > 3.0$ were
excluded from the calculation of the composite. This procedure was
repeated four times to produce the final composite.  The number of
spectra rejected at each iteration was 49, 4, 2, 0 confirming both the
stability of the procedure and the homogeneity of the bulk, $\sim
90\%$, of the galaxy spectra.  The resulting composite spectrum is
shown in Figure \ref{mspec}. Figure \ref{chi_four} displays the final
$\chi^2$ histogram of all spectra contributing to the composite. The
narrow distribution of values with a peak at $\chi^2 = 1.2$ confirms
the very high degree of homogeneity of the majority of spectra in the
early--type galaxy sample.

\begin{figure}
\psfig{figure=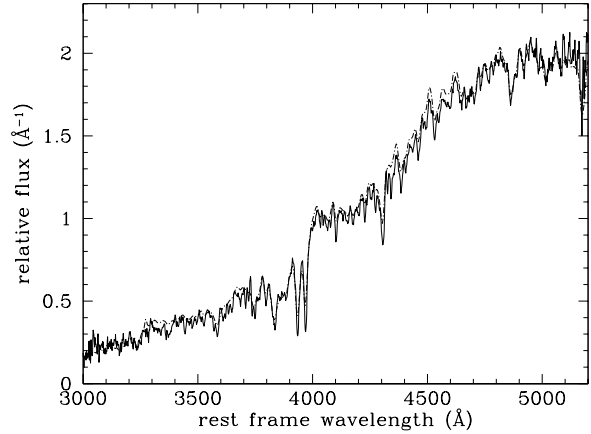,width=8.0cm,angle=270.0}
\caption{Comparison of the composite early--type galaxy spectrum
(solid line) to the Kinney template (dashed line).}
\label{compare_mspec_kin}
\end{figure}

Figure \ref{compare_mspec_kin} indicates that the composite
early--type galaxy spectrum is virtually indistinguishable from the
Kinney early--type galaxy template. The detailed resemblance between
the two spectra is not a result of the flux calibration procedure
(Section \ref{subsec_response}), which effectively matches the two
spectra only over wavelength scales $\ga 1000${\AA}. The excellent
match between the composite stellar spectrum and a template of
spectral type M2V (Figure 5) confirms the similarity of the composite
galaxy spectra to the Kinney template is not a result of the boostrap
calibration procedure. A quantitative comparison of the spectroscopic
properties of the early--type galaxy sample and the Kinney template
will presented in Paper II (Willis et al. 2001, in preparation).

\subsection[]{The variation in spectroscopic properties}

\begin{figure}
\psfig{figure=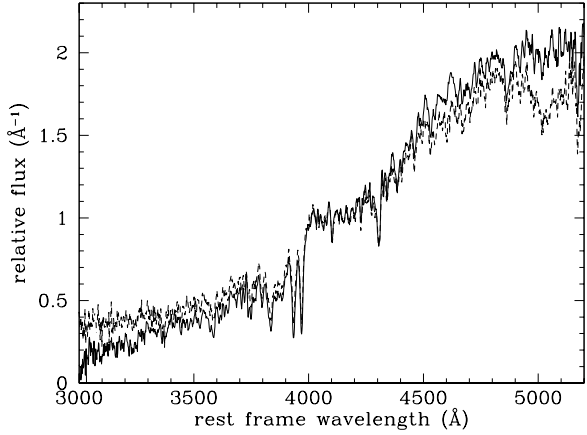,width=8.0cm,angle=270.0}
\caption{The composite spectrum (solid line) and the mean spectrum of
the 55 excluded spectra (dashed line).}
\label{c_rej}
\end{figure}

The $\sim 10${\%} (55 objects) of spectra excluded by the $\chi^2$
criterion in the generation of the composite spectrum constitute
themselves a relatively homogeneous population with objects occurring
uniformly over all fields and among all fibres within fields. Comparison of
the composite spectrum with the mean of the 55 excluded spectra (Figure
\ref{c_rej}) confirms that the spectra of the excluded objects are also
early--type galaxies but with a slightly bluer continuum slope and
marginally weaker absorption lines. Do these subtle differences reflect
intrinsic differences in the spectral energy distributions among the
target population or could they arise from limitations in the
observations or reductions?

The dominant uncertainty in the overall shape of the individual spectra
is the uncertainty associated with the vertical normalisation and shape
of the sky residual. Examination of the absorption line properties and
their relation to the variation in the shapes of the spectra offers the
prospect of discriminating between instrumental and intrinsic
explanations for the variation in spectroscopic properties.

\begin{figure}
\psfig{figure=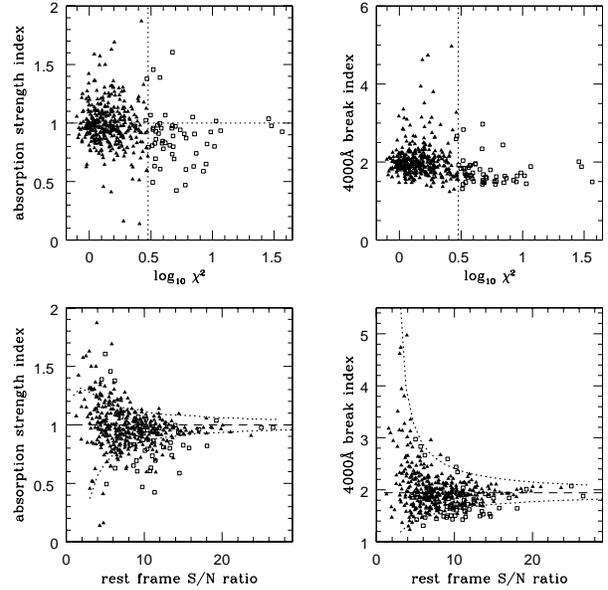,width=8.5cm}
\caption{Continuum and absorption line
properties of the early--type galaxy sample. Top left: Absorption line
strength index versus $\log_{10}\chi^2$. The vertical dotted line
indicates $\chi^2=3$. The horizontal dotted line indicates absorption
line strength index = 1.  Top right: 4000{\AA} break index versus
$\log_{10}\chi^2$. Bottom left:  Absorption line strength index versus
spectrum S/N. Bottom right:  4000{\AA} break index versus spectrum S/N.
Dotted lines in the bottom panels represent the 2$\sigma$ error in the
parameters associated with typical uncertainties in sky--subtraction.
The central dashed lines show the index associated with the composite
spectrum.}
\label{4_abs}
\end{figure}

Figure \ref{4_abs} shows the distribution of absorption line strengths
for all 433 spectra in Sample A with objects included in (378 objects)
and excluded from (55 objects) the composite identified
separately. The top left panel shows absorption line strength index
versus $\chi^2$ (at the final iteration of the composite spectrum
calculation). The absorption line strength index is the ratio of the
combined absorption line strength of the most prominent absorption
features in a spectrum to the combined absorption line strength in the
composite spectrum. A value of unity indicates the strengths are the
same and an index greater/less than unity indicates the line strength
is greater/less than in the composite spectrum (see Appendix
for an expanded discussion).  The top right panel of
Figure \ref{4_abs} shows the strength of the 4000{\AA} break feature
(D4000) in each early--type galaxy spectrum. This index is the ratio
of the the mean flux (rejecting outliers) over the
wavelength range (4030--4180{\AA} to that in the wavelength range
3750--3900{\AA}. More positive values of D4000 indicate a larger
continuum discontinuity at 4000{\AA} and vice versa.

The predicted variation in absorption line and 4000{\AA} break
indicies due to the $2${\%} uncertainty in sky--subtraction was
obtained via simulations based on the composite spectrum.  A series of
spectral templates corresponding to the composite spectrum observed at
S/N of 2 to 30, spanning the distribution in the observed sample, were
generated. An additional spectral component, corresponding to a
$2\sigma$ error in sky--subtraction was then added, or subtracted. The
absorption line strength and D4000 indices were calculated for the
simulated spectra and the results plotted as the dotted lines in
Figure \ref{4_abs} (lower plots).

The observed spread in properties of spectra included in the
calculation of the composite in Figure \ref{4_abs} is entirely
consistent with the predictions based on the instrumental/reduction
uncertainties. By contrast the 55 excluded spectra lie well away from
the envelope defined by the $\pm 2\sigma$ errors in sky--subtraction
and there is a marked excess of spectra that possess weaker absorption
and shallower D4000 breaks. Their distribution in Figure \ref{4_abs}
does not suggest they represent the tail of the distribution describing
the bulk of the galaxy spectra. The trend for spectra with high S/N to
be excluded from the composite when they possess absorption line
strength and D4000 indices indicating only small differences from the
mean arises due to the nature of the threshold $\chi^2$ criterion.

\begin{figure}
\psfig{figure=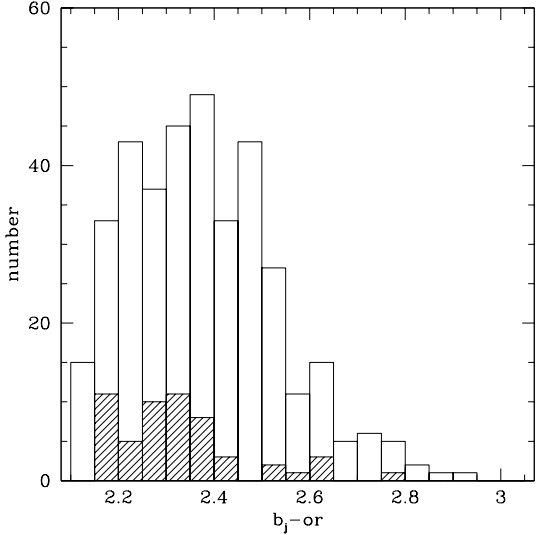,width=7.5cm,angle=0.0}
\caption{$b_j-or$ histograms of
spectra included (unfilled) and excluded (filled) from the
composite.}
\label{colour_rej_hist}
\end{figure}

Thus, examination of the inter--relation of spectroscopic properties
and the effects of instrumental/reduction uncertainties suggests the
sample of excluded spectra represent a sub--sample of objects with
bluer continua and weaker absorption features. Support for this
conclusion is offered by the distribution of $b_j - or$ colours for the
two samples of objects (Figure \ref{colour_rej_hist}). An increasing
fraction of objects with bluer observed $b_j-or$ colours is made up of
objects excluded from the composite spectrum. A Kolmogorov--Smirnov
test indicates that the two samples differ at the 92{\%} confidence
level. Taken alone, the increasing incidence of blue $b_j-or$ colours
among the sample of spectra excluded from the composite is not
significant. However, the result, when combined with the (completely
independent) differences observed in the strength of the D4000 index
from the spectra, supports the identification of the group of spectra
excluded from the composite as possessing small but distinct
differences in their spectral energy distributions compared to the
composite spectrum.

In summary, $\sim 90\%$ of the sample possess spectra essentially
indistinguishable from a redshift $z=0$ early--type galaxy spectral
energy distribution. Some $10\%$ of the sample also possess spectra
very similar to a $z=0$ early--type galaxy spectrum but absorption
line strengths are weaker by $\sim 20${\%} and the spectra are bluer
by $\sim 25${\%} across the $4000\,$\AA--break. The significance of
this sub--population in the sample is discussed in Paper II (Willis et
al. 2001, in preparation).

\subsection[]{[OII] 3727 emission}

Evidence for the presence of ongoing star formation or active galactic
nuclei in the galaxies was sought via a search for [OII] 3727 emission
in the ``continuum subtracted'' spectra of the galaxies. One goal of
the project is to identify the presence of emission lines, due to
background lensed sources, appearing at arbitrary wavelengths in the
spectra of the galaxies. As a
consequence, the continuum--subtraction algorithm employed was designed
to remove both the continuum and absorption--line signature of the
early--type galaxy over the full extent of the spectrum, leaving only
potential emission lines. The composite galaxy spectrum (Section
\ref{sec_comp_spec}) provides a natural template for removal of the
galaxy continuum and absorption line signature. However, weak [OII]
3727 emission is visible in the composite spectrum. Thus, the
wavelength region 3720-3734{\AA} was replaced by the corresponding
region of the Kinney et al. (1996) early--type galaxy SED, suitably
scaled to match the composite spectrum. Each individual galaxy spectrum
was then divided by a suitably redshifted version of the modified
composite spectrum. The resulting transformation array was median
filtered, using a 31--pixel window ($\sim 70${\AA}), in order to remove
small scale features (which might include emission lines). The modified
composite spectrum was then scaled using the median--filtered
transformation array and subtracted from the galaxy spectrum.

This procedure proved extremely effective in removing the continuum
signature of the galaxies but residual signal due to variations in the
strength of absorption lines from galaxy to galaxy was still evident.
An absorption line template, based on the presence of absorption
features visible in the composite spectrum (Table \ref{abs_table}), was
generated. A least--squares procedure was then employed to determine
the fractional scaling of the absorption line template that, when
subtracted from each residual galaxy spectrum, minimised the absorption
line signal (Appendix). The continuum and absorption line subtraction
procedures applied to each galaxy spectrum produced a residual spectrum
with zero mean flux and deviations about zero due to noise plus
excursions from any emission lines present. The noise properties are
consistent with the Poisson noise per pixel determined by propagating
the photon statistics of the original exposures through the reduction
and analysis pipeline.
 
Emission lines could then be identified in the residual spectra using
standard matched--filter techniques (e.g. Hewett et al. 1985). A
Gaussian filter of 10{\AA} FWHM, truncated at 4 FWHM, was employed as
the line template (chosen to be sensitive to the range of [OII] 3727 FWHM
anticipated from the sample of luminous early--type galaxies given the
instrumental resolution of the spectrograph). Cross--correlation of
the template with the residual galaxy spectra produced, for each
wavelength increment, a noise--weighted S/N quantifying the
significance of any emission feature, a $\chi^{2}$--value describing
the match to the shape of the Gaussian template and a vertical scaling
for the template, from which the line flux and equivalent width may be
calculated.

\begin{figure}
\psfig{figure=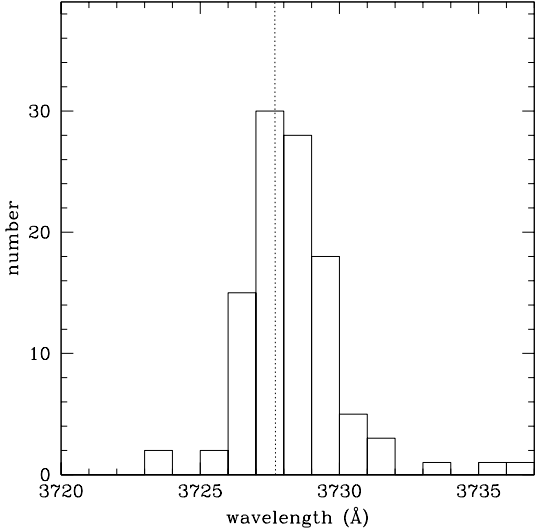,width=7.5cm,angle=0.0}
\caption{Wavelength centroid histogram for the 104 [OII] 3727
emission line detections. The vertical dashed line indicates
3727.7{\AA}, the central wavelength of the [OII] doublet.}
\label{OII_centroid}
\end{figure}

The search for [OII] 3727 emission was undertaken over the rest--frame
wavelength interval 3717{\AA} $< \lambda <$ 3737{\AA} while the two
immediately adjacent 20{\AA} intervals provided an empirical control
region. After some experimentation, candidate emission features were
selected to possess ${\rm{S/N}} \geq 3.5$ and $\chi^2 \leq 2.0$. The
central wavelengths, FWHM and flux in the candidate lines were then
calculated via a least squares fit of a three--parameter Gaussian
profile to the data. Properties of the 104 emission lines selected at
${\rm{S/N}} \geq 3.5$ and $\chi^2 \leq 2.0$ within the specified
rest--frame wavelength interval, with fluxes derived from the
three--parameter model fit, are included in Table \ref{spec_cat1}. Figure
\ref{OII_centroid} displays the number of detected emission--lines as
a function of wavelength. The reality of the [OII] emission line detections is
confirmed by the statistics of ``detections'' from the adjacent
control regions, where, from a wavelength region twice that of the
[OII] window, only 2 features satisfy the emission line selection
parameters, i.e. $\sim 1$ of the 104 emission line detections is
expected to be spurious.

\subsection[]{The number--redshift distribution and absolute magnitude
sampling}

\begin{figure}
\psfig{figure=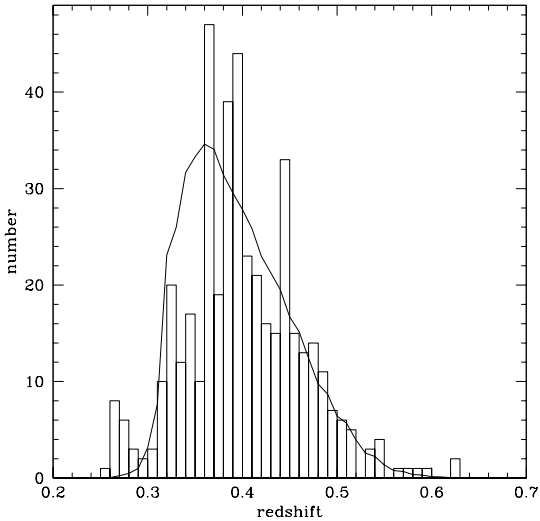,width=7.5cm}
\caption{Number--redshift histogram of 485 early--type galaxies. The
solid line indicates the number--redshift distribution predicted from
applying the survey selection criteria to a simulated, passively
evolving early--type galaxy population (see text).}

\label{nz_all}
\end{figure}

The number--redshift (N($z$)) distribution of the 485 early--type galaxies
from the 1998/05/16--17 sample is displayed in Figure \ref{nz_all}. The
galaxies span the redshift interval $0.25 \le z \le 0.63$ and the
median redshift of the sample is $z_{med}=0.391$. The observed N($z$)
distribution is consistent with the predicted distribution generated by
applying the survey selection criteria to an early--type galaxy
population described by a representative spectral evolution model and
luminosity function. A detailed discussion of the luminosity function
and space density of the population is defered until a later paper.
However, for illustration, the predicted N($z$) for the sample
employing the early--type galaxy luminosity function and evolution
parameters of Pozzetti et al. (1996) in a cosmology specified by the
parameters $\Omega=0.3$, $\Lambda=0.7$ and $H_0 = 70$ kms$^{-1}$
Mpc$^{-1}$ is also shown in Figure \ref{nz_all}.

\begin{figure}
\psfig{figure=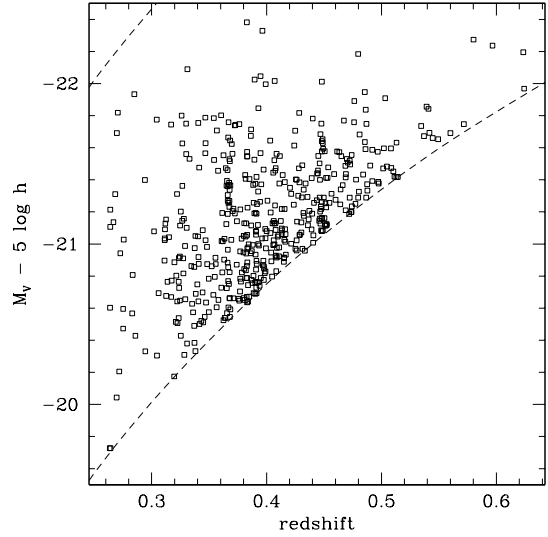,width=7.5cm}
\caption{Absolute V--magnitude versus redshift of 485 early--type
galaxies from the 1998/09/16--17 sample. The dashed lines indicate the
apparent magnitude selection criteria $16.4 < i < 18.85$ transformed
to rest--frame absolute V--magnitude according to equation
\ref{eqn_abs}.}
\label{abs_z}
\end{figure}

Figure \ref{abs_z} shows the absolute $V$--band magnitude distribution
versus redshift of the 485 early--type galaxies. Absolute magnitudes
are expressed in the $V$--band as, at the median redshift of the
sample, the UKST $i$--band samples the rest--frame $V$--band in each
early--type galaxy. Absolute $V$--band magnitudes were determined from
apparent $i$--band magnitudes and redshifts according to

\begin{equation}
{
M_V(z) = m_i - 25 - 5 \log d_L(z) - e_i(z) - k_i(z) + (V-i)_0,
}
\label{eqn_abs}
\end{equation}

{\noindent}where $m_i$ is the apparent $i$--band magnitude of a given
galaxy at a redshift $z$, $d_L(z)$ is the luminosity distance at a
redshift $z$ within the specified cosmological model, $e_i(z)$
parameterizes the luminosity evolution in the $i$--band, $k_i(z)$ is
the $k$--correction for the $i$--band and $(V-i)_0$ is the $V-i$ colour
for an early--type galaxy at $z=0$. The dashed lines in Figure
\ref{abs_z} indicate the absolute magnitude limits of the sample as a
function of redshift defined by the $i$--band apparent magnitude
selection criteria (Table \ref{tab_photo_select}), confirming that the
majority of the early--type galaxy sample consists of highly--luminous
galaxies ($ -22.4 \la M_V \la -19.5$).

\subsection[]{Redshift catalogue}

Positions and magnitudes (Section 2), redshifts (Section 3.5) and
[OII] 3727 emission line fluxes (Section 4.3) for the 485 galaxies
observed during the 1998 September observations are presented in Table
\ref{spec_cat1}. The same information, aside from the emission line
fluxes which are not available, for an additional 96 galaxies observed
during 2dF runs prior to 1998 September are presented in Table
\ref{spec_cat2}. The complete redshift catalogue can be accessed via
the Strasbourg astrophysical Data Center (CDS) located at the URL: {\tt
http://cdsweb.u-strasbg.fr/Cats.html}.

\begin{table*}
  \vbox to220mm{\vfil Landscape table to go here.
  \caption{}
 \vfil}
 \label{spec_cat1}
\end{table*}

\begin{table*}
%
%
\begin{minipage}{140mm}
\caption{Spectroscopic catalogue of 96 early--type galaxies based on
pre--1998 September 2dF observations.}
\label{spec_cat2}
\begin{tabular}{ccccccccccc}
Galaxy & {RA\footnote{Positions are J2000.0.}} & {Dec.} &
$b_j$ & $or$ & $i$ & error($b_j$) & error($or$) & error($i$) & $z$ & $cz_{err}$\footnote{A zero value indicates
redshift assigned visually.} \\
&(hh:mm:ss.ss)&(dd:mm:ss.s)&&&&&&&&(kms$^{-1}$) \\
&&&&&&&&&& \\
G1344m0020 & 13:44:55.61 & -00:20:16.0 & 20.94 & 18.86 & 18.38 & 0.11 & 0.07 & 0.11 & 0.417 &  70 \\
G1343m0023 & 13:43:31.45 & -00:23:57.6 & 19.78 & 18.18 & 17.48 & 0.07 & 0.05 & 0.06 & 0.330 &  51 \\
G1342m0027 & 13:42:44.75 & -00:27:32.3 & 20.16 & 18.86 & 18.00 & 0.09 & 0.07 & 0.09 & 0.327 & 128 \\
G1345m0019 & 13:45:00.77 & -00:19:54.4 & 20.92 & 19.13 & 18.45 & 0.12 & 0.08 & 0.11 & 0.382 &  81 \\
G1342m0026 & 13:42:37.78 & -00:26:53.6 & 20.99 & 19.63 & 18.68 & 0.15 & 0.10 & 0.13 & 0.421 &   0 \\
G1342m0021 & 13:42:59.67 & -00:21:10.5 & 20.73 & 18.76 & 17.95 & 0.12 & 0.07 & 0.08 & 0.380 &  77 \\
G1342m0030 & 13:42:39.99 & -00:30:51.8 & 20.07 & 18.59 & 17.68 & 0.09 & 0.06 & 0.07 & 0.326 &  91 \\
G1342m0038 & 13:42:05.46 & -00:38:29.0 & 20.65 & 18.74 & 18.16 & 0.10 & 0.07 & 0.10 & 0.379 &  81 \\
G1343m0043 & 13:43:11.57 & -00:43:29.8 & 20.74 & 19.09 & 18.34 & 0.12 & 0.08 & 0.11 & 0.401 & 147 \\
G1342m0044 & 13:42:27.18 & -00:44:19.6 & 20.78 & 19.00 & 18.41 & 0.11 & 0.08 & 0.11 & 0.381 &  86 \\
G1342m0019 & 13:42:05.73 & -00:19:07.7 & 20.05 & 18.57 & 17.81 & 0.08 & 0.06 & 0.08 & 0.329 &  43 \\
G1342m0100 & 13:42:56.40 & -01:00:22.2 & 21.25 & 19.36 & 18.61 & 0.16 & 0.09 & 0.12 & 0.386 &  85 \\
G1342m0014 & 13:42:57.05 & -00:14:24.1 & 21.12 & 19.04 & 18.25 & 0.15 & 0.08 & 0.10 & 0.396 &  71 \\
G1341m0050 & 13:41:56.26 & -00:50:43.8 & 20.93 & 19.42 & 18.66 & 0.13 & 0.09 & 0.13 & 0.345 & 121 \\
G1342m0103 & 13:42:26.40 & -01:03:33.5 & 20.44 & 18.90 & 18.00 & 0.10 & 0.07 & 0.09 & 0.391 &  64 \\
G1341m0019 & 13:41:03.63 & -00:19:42.3 & 20.58 & 19.17 & 18.38 & 0.11 & 0.08 & 0.11 & 0.340 &  74 \\
G1341m0023 & 13:41:13.76 & -00:23:08.5 & 20.53 & 19.21 & 18.26 & 0.11 & 0.08 & 0.10 & 0.338 &  46 \\
G1340m0108 & 13:40:58.38 & -01:08:39.2 & 19.10 & 17.71 & 16.87 & 0.05 & 0.04 & 0.05 & 0.294 &  68 \\
G1341m0037 & 13:41:11.69 & -00:37:05.1 & 20.99 & 19.06 & 18.26 & 0.14 & 0.08 & 0.10 & 0.356 &  67 \\
G1341m0058 & 13:41:21.72 & -00:58:50.4 & 20.94 & 19.03 & 18.45 & 0.12 & 0.08 & 0.11 & 0.384 &  55 \\
G1340m0101 & 13:40:52.86 & -01:01:55.7 & 21.13 & 19.32 & 18.65 & 0.14 & 0.09 & 0.13 & 0.416 & 150 \\
G1340m0058 & 13:40:41.52 & -00:58:08.7 & 21.06 & 19.34 & 18.62 & 0.14 & 0.09 & 0.12 & 0.355 & 134 \\

\end{tabular}								  
\end{minipage}								  
\end{table*}								 
	
\section{Summary}
	
Luminous, field early--type galaxies at cosmologically significant
distances represent an important fraction of the population of normal
galaxies that is almost entirely absent from the current generation of
spectroscopic surveys.  Identification of such galaxies out to
redshifts of $z \sim 0.6$ can be readily achieved by applying a
straightforward $BRI$ broadband colour selection to photographic or CCD
material that reaches only modest depth, $I \simeq 19$. Morphological
information allows contaminating M dwarfs to be eliminated as
candidates. However, for magnitudes $17 \le I \le 19$, and given
photometric precision of $\la 0.15\,$mag, the number of galaxies
greatly exceeds the number of stellar objects and stellar contamination
is not a significant limitation in compiling galaxy samples. The key
requirement for the assembly of a large sample of such galaxies is
sky coverage as the surface density of objects to $I < 19.0$ is 
$< 100\,$deg$^{-2}$.

Using APM scans of UKST $b_j$ $or$ and $i$ photographic plates we have
compiled a photometric catalogue of 9599 candidate galaxies,
magnitudes $m_i \le 18.95$, over $220\,$deg$^2$ in seven
UKST fields. From the results of spectroscopic observations the
contamination by stellar objects, M dwarfs, is only $\sim 2\%$.

Spectroscopic observations of a sample of 581 galaxies from the
photometric catalogue are presented. The galaxies span the redshift
interval $0.25 \le z \le 0.63$ with absolute magnitudes $-22.4 \le M_V
\le -19.5$. The absorption and emission line properties of the sample are
parameterised and a high signal--to--noise ratio composite spectrum is
constructed. The galaxy spectra display very similar spectral energy
distributions, including a narrow range of absorption line strengths
and 4000{\AA}--break indices. Comparison of the spectra with that of a
local early--type galaxy template demonstrates that the rest--frame
optical spectra of the galaxies in the sample are consistent with an
old, passively evolving stellar population with little evidence of
recent star--formation.

The sample represents a unique resource  and will form the basis for 
investigations in the fields of gravitational lensing, galaxy evolution
as a function of environment and the evolution of large scale structure
to be presented in forthcoming papers.

\section*{Acknowledgments}
The staff of the Anglo--Australian Observatory provided invaluable
assistance during the acquisition of the spectroscopic observations.
We are grateful for the active longterm support of of the United
Kingdom Schmidt Telescope Unit and the staff of the Automated Plate
Measuring facility. The Isaac Newton Telescope is operated on the
island of La Palma by the Isaac Newton Group in the Spanish
Observatorio del Roque de los Muchachos of the Instituto de
Astrofisica de Canarias. JPW acknowledges the support of a PPARC
research studentship and financial support from IoA, Cambridge.  The
project would not have been possible without the data and analysis
facilities provided by the Starlink Project which is run by CCLRC on
behalf of PPARC.

\appendix

\section[]{Absorption line strength}
\label{app}

The 2dF spectra of the galaxy sample possess relatively low S/N and are
of only intermediate, $\sim 4$\AA, resolution. Equivalent widths of
individual absorption features are thus very uncertain. To provide a
more reliable measure of absorption line strength for each galaxy an
absorption line template was fitted to all prominent absorption
features in each spectrum, thus maximising the S/N of a measurement of
absorption line strength.  The combined absorption line strength in a
particular early--type galaxy spectrum is expressed as a fraction of
the absorption line strength in the composite galaxy spectrum.

\begin{figure}
\psfig{figure=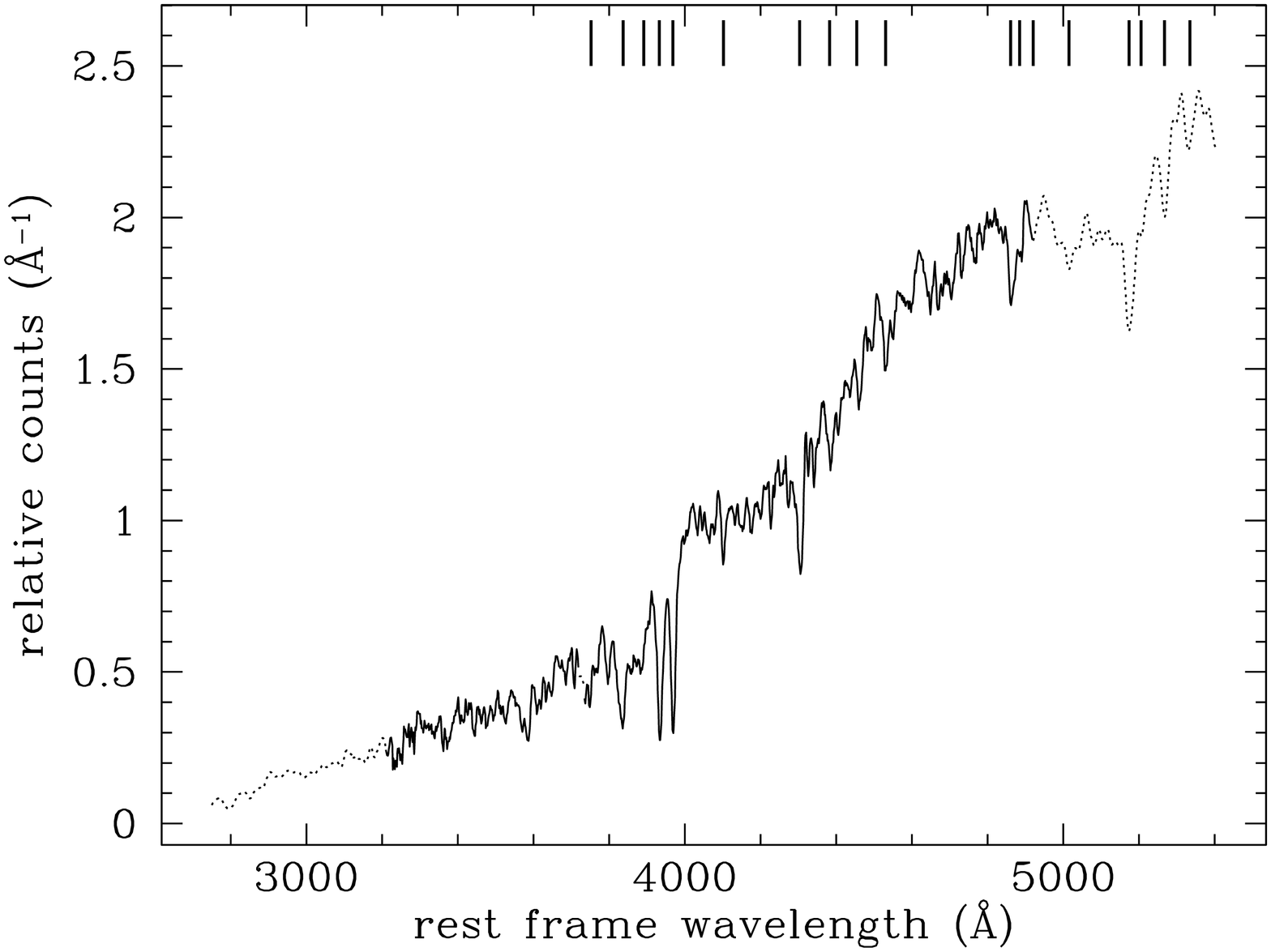,width=8.0cm,angle=270.0}
\caption{Composite early--type galaxy spectrum employed in absorption
line analyses. The mean early--type galaxy spectrum is indicated by
the solid line and the Kinney early--type template by the dotted
line. Vertical solid lines mark the wavelength centres of absorption
features listed in Table \ref{abs_table}.}
\label{mspec_comp}
\end{figure}

Rest--frame wavelength regions associated with significant absorption
features were identified by selecting features in the composite galaxy
spectrum displaying a flux decrement greater than 12.5{\%} relative to
the locally--determined continuum. Absorption features identified via
this method are listed in Table \ref{abs_table} and are indicated in
Figure \ref{mspec_comp}.

\begin{table}
\caption{Absorption features identified in the composite spectrum.}
\label{abs_table}
\begin{tabular}{lcc}
Feature & Wavelength ({\AA}) & Interval ({\AA}) \\ 
H$\kappa$ & 3752.2 &  3744.5--3753.5 \\ 
H$\eta$   & 3837.5 &  3822.5--3896.0 \\ 
H$\zeta$  & 3891.2 &  3822.5--3896.0 \\ 
CaII K    & 3933.4 &  3924.5--3952.0 \\ 
CaII H    & 3969.2 &  3952.5--3980.0 \\ 
H$\delta$ & 4102.8 &  4097.0--4109.0 \\ 
G--band    & 4303.4 &  4295.0--4316.0 \\
Fe4383    & 4383.0 &  4380.5--4392.5 \\ 
Ca4455    & 4455.0 &  4455.5--4464.5 \\ 
Fe4531    & 4531.0 &  4526.0--4536.5 \\ 
H$\beta$  & 4861.3 &  4853.0--4880.0 \\ 
Fe4890    & 4890.0 &  4884.5--4895.0 \\ 
blend     & 4921.0 &  4917.5--4925.0 \\ 
Fe5015    & 5015.0 &  5012.0--5024.0 \\ 
Mgb       & 5174.0 &  5147.0--5196.5 \\ 
blend     & 5206.0 &  5201.0--5214.5 \\ 
Fe5268    & 5268.0 &  5256.5--5280.5 \\ 
Fe5335    & 5335.0 &  5324.0--5346.5 \\
\end{tabular}
\end{table}

To prepare the galaxy spectra for the absorption line strength
measurement it is necessary to define a reliable continuum.  Each
early--type galaxy spectrum was divided by a suitably scaled and
redshifted version of the composite galaxy spectrum to generate an
exact transformation between the spectra. The transformation array was
median--filtered with a 31--pixel window ($\sim 70${\AA}) to remove
positive and negative deviations on wavelength scales $\la 30${\AA} yet
preserve large--scale continuum variations. The composite spectrum was
then scaled using the median--filtered transformation array to match
each galaxy spectrum. Absorption line spectra for data and matched
template spectra were generated by fitting and subtracting continuum
profiles.

The absorption line strength measure for a galaxy is calculated from
the scaling of the absorption line template to the galaxy's absorption
line profile generated as described above. In the presence of random
noise the most probable value of the scaling factor that matches the
absorption line template to a galaxy's absorption line profile is
obtained by minimising the weighted sum of the squared error residuals
between the data and the template. If $D_i$ is the data spectrum, $P_i$
is the template absorption profile and $\sigma_i$ is the 1$\sigma$
noise per pixel, one then minimises the $\chi^2$ expression $\chi^2 =
\Sigma_i {(D_i-k P_i)}^2 / {\sigma_i}^2$ with respect to the scale
factor $k$. The sum runs over the $i$ pixels of the absorption line
template. The value of the scale factor that minimises the sum is given
by

\begin{equation}
k = \frac{\Sigma_i ({D_i}{P_i}) / {\sigma_i}^2} {\Sigma_i {(P_i /
	\sigma_i)}^2}.
\label{eqn_k}
\end{equation}

Thus a scale factor of $k$=1 means that the line strength in a
spectrum is equal to the line strength in the template
whereas $k>$1 means that the line strength is greater than the
template, and vice versa.

\bsp

\label{lastpage}

\end{document}